\documentclass[twocolumn,aps,showpacs,pra]{revtex4}
\usepackage{epsfig}
\usepackage[english]{babel}
\usepackage{latexsym}
\usepackage{graphics}
\usepackage{subfigure}
\usepackage{epsfig}
\usepackage{graphicx}
\usepackage{dcolumn}
\usepackage{amsmath}

\begin{document}
\title{High-order harmonic generation from
diatomic molecules with large internuclear distance: The effect of
two-center interference}
\author{Y. J. Chen$^{1}$,  J. Liu$^{1*}$}

\date{\today}

\begin{abstract}
In the present paper, we investigate  the high-order harmonic
generation (HHG) from diatomic molecules with  large internuclear
distance using a strong field approximation (SFA) model. We find
that the hump and dip structure emerges in the plateau region of
the harmonic spectrum, and  the location of this striking
structure  is sensitive to the laser intensity. Our model analysis
reveals that two-center interference as well as  the interference
between different recombination electron trajectories are
responsible for the unusual enhanced or suppressed harmonic yield
at a certain order, and these interference effects are greatly
influenced by the laser parameters such as intensity.
\end{abstract}
\affiliation{1.Institute of Applied Physics and Computational
Mathematics, P.O.Box 100088, Beijing, P. R. China\\
2.Graduate School, China Academy of Engineering Physics, P.O. Box
8009-30, Beijing, 100088, P. R. China} 
\maketitle
\section{Introduction}
The high-order harmonic generation (HHG) from  atoms and molecules
has been one of the most intensely studied aspects of strong-field
physics\cite{lein1,lein2,lein3,Bandrauk,Ahn,Nalda,c.vozzi,Tsuneto,Dietrich,X.
X. Zhou,J.Itatani,Charge-resonance,Ciappina,cj,J}, because it can
be applied as a coherent ultrashort radiation source in the
extreme ultraviolet(XUV) and soft x-ray regions\cite{M.
Hentschel}. Due to the multi-center characteristics and greater
freedom, the molecules show more complicated structures in their
HHG spectrum than atoms and the dynamics of molecules in the
external field is more difficult to investigate.

Among many treatments, the semiclassical recollision
model\cite{Corkum} is an applaudable theory that provides a
appropriate picture for HHG. In this model, the high harmonics are
generated by a three-step sequence: (1) tunnel ionization of the
highest energy of the electron, (2) acceleration of the free
electron in the laser field, and driving the electron back to the
parent ion, (3) recombination of the electron to the state from
which it originated. The recombination step leads to the emission
of a XUV photon whose energy is given by the sum of the electron
of kinetic energy plus the ionization potential of the state. A
widely used semianalytical approach to describe HHG is the strong
field approximation  (SFA) or Lewenstein model \cite{Lewenstein},
which is applicable to the  molecules. It can be regarded as the
quantum mechanical version of the semiclassical recollision model.
The Lewenstein model has the advantage of requiring less
computational effort than the $ab\ initio$ solution of the
time-dependent Schr\"{o}dinger equation, which becomes very
demanding at high laser intensities. Another advantage of the
model is that it yields a physical interpretation of the
underlying mechanisms and a certain degree of analytical
description.

Recently, several theoretical and experimental reports study
two-center interference on the HHG from molecules of
H$_2^+$\cite{lein1,lein2,lein3,Bandrauk} and
CO$_2$\cite{Ahn,Nalda,c.vozzi,Tsuneto}. Lein $et\ al$ observed the
phenomenon and attributed it to the returning electron wave
colliding with cores of the two-center molecule, where an
interference occurs\cite{lein1}. They  found that the interference
effects are sensitive to the molecular orientation but not to the
laser parameters.  Ciappina $et\ al$ use the SFA model to study
two-center interference\cite{Ciappina}. They compared the SFA
prediction of the interference-minimum position to the exact value
obtained from the $ab\ initio$ calculations. They observe that
using the two-center continuum wave functions, instead of a plane
wave to describe the continuum electron, greatly improves the
comparison. The usage of the continuum wave functions can be
considered  an attempt to account for Coulomb effects on the
returning electron.

The previous  discussions, however, are mainly toward the
molecules with small internuclear distance $R$. Detailed
investigations for  the interference effects on HHG from molecules
with large  $R$ are still in lack. For the large internuclear
distance case, strong coupling between the ground state and the
first excited state emerges due to  charge-resonance
effect\cite{Charge-resonance}. This could  complicate the HHG
process.

In the present paper, using a  SFA model that considers the
Coulomb effects, we investigate the HHG from a 1D model diatomic
molecule with large $R$ ($16$ a.u.). The SFA model calculation
gives the HHG spectrum qualitatively consistent with the numerical
result and shows the pronounced hump and dip structure in the
plateau region. Detailed analysis shows that  two-center
interference as well as  the interference between different
recombination electron trajectories are responsible for the
enhanced or suppressed harmonic yield at a certain order. And in
particular, the location of the unusual structure is found to be
sensitive to the laser intensity.

This paper is organized as follows. In Sec.II, we present our
analytical theory. In Sec.III, we apply our theory to the HHG from
a 1D model diatomic molecule with large $R$.
We also analyze the complicated interference patterns
here. Sec.IV is our conclusion.

\section{Analytic theory}
In Ref.\cite{Charge-resonance}, we  have developed a SFA model for
the HHG from symmetric diatomic molecules, emphasizing the
influence of the charge-resonance  states that are strongly
coupled to electromagnetic fields for the case of large $R$. And
we showed there that for sufficiently large internuclear
distances, while initially the system is in the ground state, the
contribution to the harmonic  comes mostly from the continuum
state-ground state transition. For very large $R$ with
$\mathbf{D}\approx 0$, where
$\mathbf{D}=\langle0|\hat{\mathbf{p}}|1\rangle$, $|0\rangle$ and
$|1\rangle$ are the ground state and the first excited state of
the unperturbed system, respectively, the harmonic formula along
the field direction can be written as
\begin{eqnarray}
\begin{split}
&P({\omega}')=i\int d\mathbf{p}|{\mathbf{A}_{0}}|\pi\langle
0|\mathbf{p}\rangle (\mathbf{p}\cdot\hat{\mathbf{e}})^{2}\langle
\mathbf{p}|0\rangle\sum_{n,m=-\infty}^{n,m=+\infty}\\
&\bigg\{\frac{\delta[(m-1+n){\omega}-{\omega}']}{A}+\frac{\delta[(m+1+n){\omega}-{\omega}']}{B}\bigg\}J_{n}J_{m},\\
\end{split}
\end{eqnarray}
where$J_{n}=J_{n}(\frac{-\mathbf{P}\cdot{\mathbf{A}_{0}}}{{\omega}}),$$J_{m}=J_{m}(\frac{\mathbf{P}\cdot{\mathbf{A}_{0}}}{{\omega}})$,
 $A=(m-1){\omega}-\frac{\mathbf{p}^{2}}{2}-E_{0},$
 $B=(m+1){\omega}-\frac{\mathbf{p}^{2}}{2}-E_{0}.$ $E_{0}$ is the
ionization potential of the ground state,
$\mathbf{A}_{0}=E\hat{\mathbf{e}}/\omega$.  $E$ is the field
amplitude, $\hat{\mathbf{e}}$ is the unit vector along the field
direction, $\omega$ is the field frequency. Eq. (1)  is applicable
for the large internuclear distance and the weak field as
discussed in our previous paper.  In the SFA\cite{Lewenstein}, an
important assumption is that electrons in the continuum states can
be treated as a free particle moving in the electric field without
considering the Coulomb potential. Accordingly, the continuum wave
function $|\mathbf{p}\rangle$ is approximated by a plane wave
$|\mathbf{p}\rangle=e^{i\mathbf{p}\cdot \mathbf{r}}$ with the
energy $E_{\mathbf{p}}=\mathbf{p}^{2}/2$, where the binding
potential is completely omitted. Here, we extend to  consider the
modification from the  Comloub potential, thus a better
approximation of the continuum states in the following form is
exploited,
\begin{equation} |\mathbf{p}\rangle=e^{i\mathbf{p_{k}}\cdot
\mathbf{r}},
\end{equation}
where the effective momentum
$\mathbf{p_{k}}=\mathbf{p}\sqrt{\mathbf{p}^{2}+2E_{0}}/|\mathbf{p}|$,
and the  energy $E_{\mathbf{p}}=\mathbf{p}^{2}/2$. This
incorporates some effects of the binding potential through its
dependence on $E_{0}$\cite{lein2,Bandrauk,J,Levesque}. With the
rectification of Eq. (2), the harmonic formula of Eq. (1) can be
rewritten  as
\begin{eqnarray}
\begin{split}
&P({\omega}')=i\int d\mathbf{p}|{\mathbf{A}_{0}}|\pi\langle
0|\mathbf{p_{k}}\rangle
(\mathbf{p}\cdot\hat{\mathbf{e}})^{2}\langle
\mathbf{p_{k}}|0\rangle\sum_{n,m=-\infty}^{n,m=+\infty}\\
&\bigg\{\frac{\delta[(m-1+n){\omega}-{\omega}']}{A}+\frac{\delta[(m+1+n){\omega}-{\omega}']}{B}\bigg\}J_{n}J_{m},\\
\end{split}
\end{eqnarray}

For comparison with 1D numerical simulation, the model is
simplified to an one dimensional model and the integral over the
momentum is evaluated by the so-called pole approximation\cite{d}
throughout this paper. By setting $A=0$ or $B=0$, which denotes
the energy conservation in the ionization process, one can obtain
the momentum $\mathbf{p}$ with certain $m$. The two $delta$
functions in Eq. (3) denote the emission of harmonics with certain
$m$ and $n$ in the recombination process.  From the above explicit
expression, we conclude that the recombination electrons with
diverse momenta could contribute to a common harmonic order, since
the $delta$ functions explicitly depend on the integer $n$. This
implies that, not only in the ionization process the electron can
absorb $m\pm1$ photons, but also in the recombination process the
electron  can absorb (positive) or emit (negative) $n$ photons,
which induces the final emission of $m\pm1 +n$ photons.

The integrand of Eq. (3) can be divided into several parts. The
part of  $\langle 0|\mathbf{p_{k}}\rangle
(\mathbf{p}\cdot\hat{\mathbf{e}})^{2}\langle
\mathbf{p_{k}}|0\rangle$ explicitly depends on the ionization
energy $E_0$ and the internuclear distance $R$, but is independent
of the field parameters. It incorporates the interference effects
between cores of the two-center molecules \cite{lein2,Muth-Bohm},
and mainly is determined by the properties of the molecules. The
part of $J_{n}J_{m}$ explicitly depends on the field amplitude $E$
and the field frequency $\omega$, representing the probability
amplitude for an electron to absorb or emit $n+m\pm 1$ photons. It
reflects the interaction between the electron and the field and is
closely related to the field parameters. Thus, those recombination
electrons  that have the momenta satisfying the condition of $A=0$
or $B=0$ all could contribute to a harmonic order, and the weight
of the contribution depends on not only two-center interference
but also the interference between different recombination electron
trajectories\cite{Lewenstein}. The latter is strongly depends on
the laser intensity, as a result, we expect that the interference
structure in  HHG spectra could rely on the field parameters. The
above theoretical analysis is verified by our numerical
simulations as shown in following.

\section{Numerical Results}

The Hamiltonian of the 1D model diatomic molecule studied here is
$H(t)$=$-\frac{d^{2}}{2dx^{2}}$$-\frac{Z}{\sqrt{1.44+({x+0.5R})^{2}}}$
$-\frac{Z}{\sqrt{1.44+({x-0.5R})^{2}}}$$-x\mathcal{E}\sin({\omega}t)$,
where $Z$ is the effective charge, $R$ is the internuclear
separation, $\mathcal{E}$ is the amplitude of the external
electric field, and ${\omega}$ is the frequency of the external
field. In the paper, we adopt the atomic units, $\hbar=e=m_e=1$.
Calculations have been performed for $780$ nm trapezoidally shaped
laser pulses with a total duration of $10$ optical cycles and
linear ramps of three optical cycles. Numerically, the above
Schr\"odinger equation can be solved by the operator-splitting
method\cite{Bambi}.

A typical result is presented in Fig. 1. Here we plot the harmonic
spectra  of a 1D model diatomic molecule  with $Z=1$, $R=16$ a.u.
and $E_{0}=0.638$ a.u. at the field intensity $I=5.3\times10^{13}
\mathrm{W/cm^2}$(Figs. 1(a) and (c)) and $I=1\times10^{14}
\mathrm{W/cm^2}$(Figs. 1(b) and (d)). Figs. 1(a) and (b) are the
numerical results and Figs. 1(c) and (d) are the analytic results
calculated by Eq. (3).  The  minimum or dip structure is at the
17th and the 15th order in Figs. 1(a) and (c), respectively, and
that in Figs. 1(b) and (d) both are at the 23th order, as
indicated by the vertical arrows. Though there are some
differences in the accurate positions of the minimum between the
analytic and the numerical results, the dip structure can be
distinguished in all plottings of the spectra. In addition, a hump
structure around the 11th order harmonic in Figs. 1(a), around the
15th order harmonic in Figs. 1(b), around the 9th order harmonic
in Figs. 1(c), and around the 13th order harmonic in Figs. 1(d),
can be identified, as indicated by the horizontal arrows.
\begin{figure}[tbh]
\begin{center}
\rotatebox{0}{\resizebox *{8.5cm}{8.0cm} {\includegraphics
{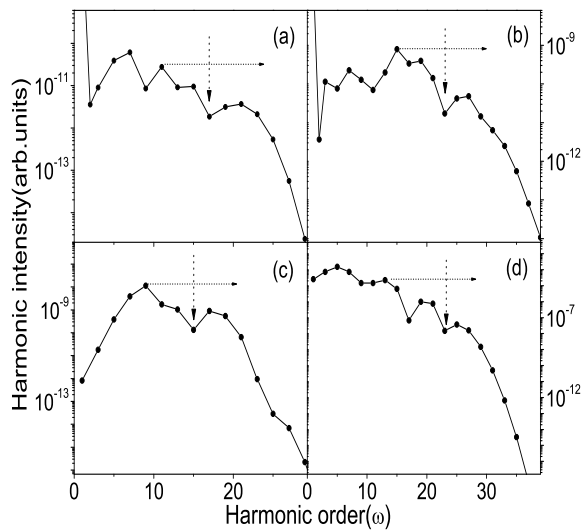}}}
\end{center}
\caption{Photon-emission spectra of a symmetrical diatomic
molecule with $Z=1$, $R=16$ a.u. and $E_{0}=0.638$ a.u. at the
field intensity $I=5.3\times10^{13} \mathrm{W/cm^2}$(Figs. 1(a)
and (c)) and $I=1\times10^{14} \mathrm{W/cm^2}$(Figs. 1(b) and
(d)). (a) and (b): The numerical results; (c) and (d): the
analytic results calculated by Eq. (3). See the context for the
illumination of the arrows.} \label{Fig. 1}
\end{figure}

\begin{figure}[tbh]
\begin{center}
\rotatebox{0}{\resizebox *{8.5cm}{8.0cm} {\includegraphics
{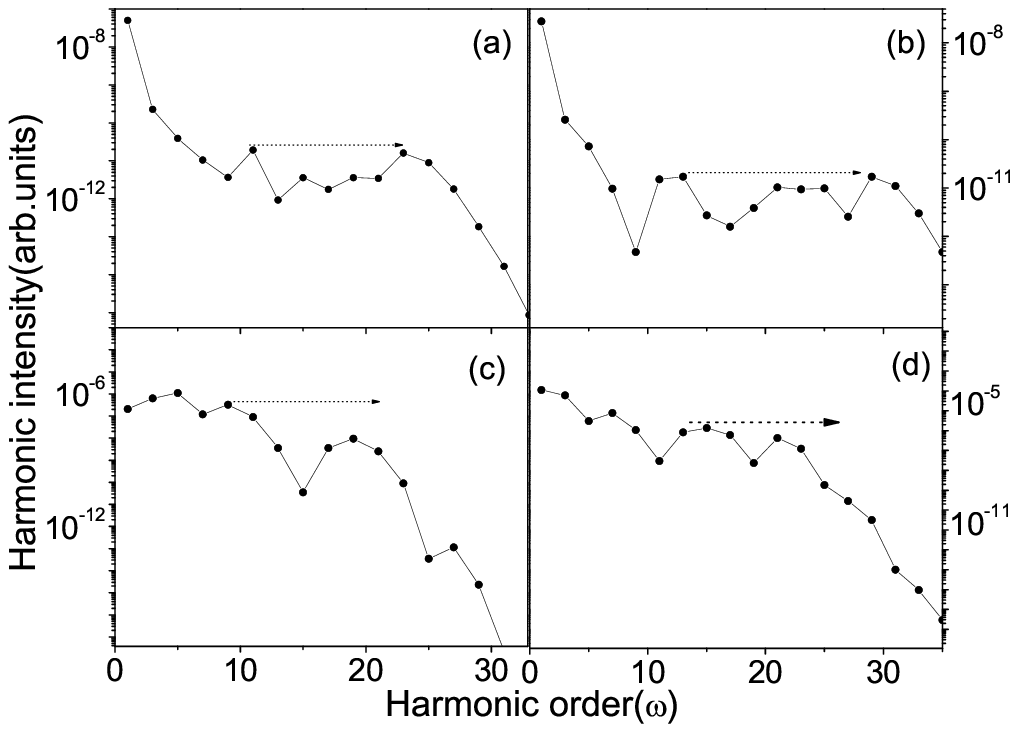}}}
\end{center}
\caption{Photon-emission spectra of a symmetrical diatomic
molecule with $Z=0.6$, $R=2.5$ a.u. and $E_{0}=0.6$ a.u. at the
field intensity $I=1\times10^{14} \mathrm{W/cm^2}$ (Figs. 1(a) and
(c)) and $I=1.5\times10^{14} \mathrm{W/cm^2}$(Figs. 1(b) and (d)).
(a) and (b): The numerical results; (c) and (d): the analytic
results calculated by Eq. (3). See the context for the
illumination of the arrows.} \label{Fig. 2}
\end{figure}

\begin{figure}[tbh]
\begin{center}
\rotatebox{0}{\resizebox *{8.5cm}{8cm} {\includegraphics
{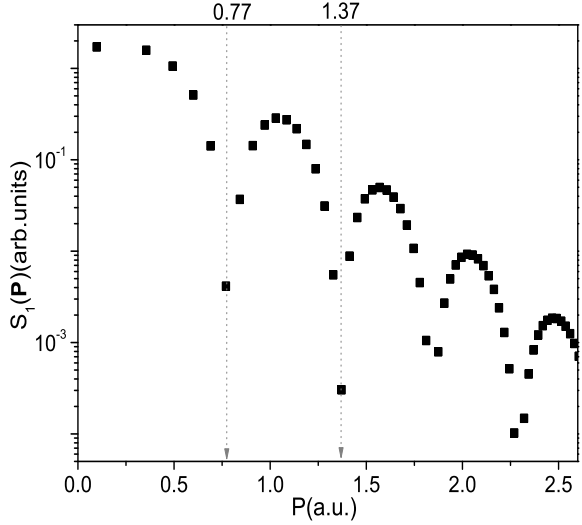}}}
\end{center}
\caption{The relationship between the function $S_{1}(\mathbf{p})$
and the momentum $\mathbf{p}$ for the chosen model molecule in
Fig. 1. See the context for the illumination of the arrows and the
numbers above the arrows.} \label{fig.3}
\end{figure}
\begin{figure}[tbh]
\begin{center}
\rotatebox{0}{\resizebox *{8.5cm}{8.0cm} {\includegraphics
{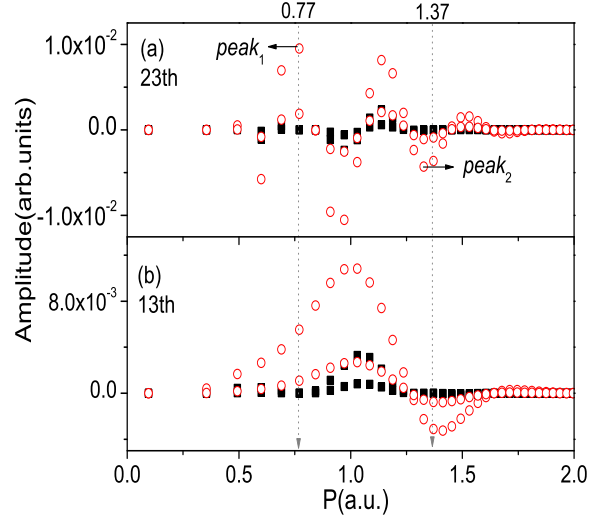}}}
\end{center}
\caption{The  values of $S_{2}(\mathbf{p})$ (the red curves with
the hollow symbol) and $S(\mathbf{p})$ (the black curves with the
solid symbol) as functions of the electron momentum $\mathbf{p}$
for certain individual harmonic orders. Harmonic orders are as
indicated. The laser and molecule parameters are the same as in
Figs. 1(b). Only the contributions from positive momenta are
shown. See the context for the illumination of the arrows and the
numbers above the arrows.} \label{fig.4}
\end{figure}

In Fig. 2, we  plot the harmonic spectra of a 1D model diatomic
molecule with $Z=0.6,R=2.5$ a.u. and $E_{0}=0.6$ a.u. at the field
intensity $I=1\times10^{14} \mathrm{W/cm^2}$ (Figs. 2(a) and (c))
and $I=1.5\times10^{14} \mathrm{W/cm^2}$(Figs. 2(b) and (d)).
Figs. 2(a) and (b) are the numerical results and Figs. 2(c) and
(d) are the analytic results calculated by Eq. (3).
The harmonic spectra in Fig. 2 show a board and flat
suppressed region, which is nearly spreading in the whole plateau
region, as indicated by the horizontal arrows. In contrast, for
the large internuclear distance cases as indicated by Fig. 1, the
harmonic spectra  show some kind of hump or dip structure.
Especially as the field intensity increases, the hump structure
become more pronounced around the 15th order harmonic in Figs.
1(b) with $I=1\times10^{14} \mathrm{W/cm^2}$, compared to that in
Figs. 1(a) with $I=5.3\times10^{13} \mathrm{W/cm^2}$. Our
calculations have been extended to the  model molecules with
$R=10$ a.u. and $R=12$ a.u. at different laser intensity. They
exhibit the similar phenomena as revealed in Fig. 1.

Fig. 3 shows the relationship of the momentum $\mathbf{p}$ and the
function
\begin{equation}
S_{1}(\mathbf{p})=\langle 0|\mathbf{p_{k}}\rangle
(\mathbf{p}\cdot\hat{\mathbf{e}})^{2}\langle
\mathbf{p_{k}}|0\rangle,
\end{equation}
for the chosen model molecule in Fig. 1. The arrows in it indicate
those momenta at which the minimal extrema of Eq. (4) appear. The
relevant momentum values are  labelled above the arrows. The
neighboring peaks and hollows in Fig. 3, corresponding to certain
maximal and minimal extrema of Eq. (4), are very close.

According to the simple picture regarding the nuclei as point
emitters proposed by Lein $et\ al$\cite{lein2}, for molecules with
symmetric initial states, destructive interference (minimum in the
harmonic spectrum) occurs when $p_{k}R\cos(\theta)=(2n+1)\pi,
n=0,1,2,...$. Here $p_{k}=|\mathbf{p_{k}}|$ is the effective
momentum of the electron, $\theta$ is the angle between the
molecular axis and the laser polarization direction. Constructive
interference (maximum in the harmonic spectrum) occurs when
$p_{k}R\cos(\theta)=2n\pi,
 n=0,1,2,...$\cite{Bandrauk,lein2}. In the 1D case, for $R=2.5$ a.u., the
first predicted minimum is at the 13th order harmonic, which
agrees with that in Figs. 2(a). While in Figs. 2(b), the minimum
shifts to the 17th order harmonic. For the large $R=16$ a.u., the
predicted minima are at the 8th order with $n=2$, the 16th order
with $n=3$, and the 27th order with $n=4$, etc. The maxima are at
the 12th order with $n=3$ and the 21th order with $n=4$, etc.
While in the numerical cases, the pronounced minimum is at the
17th order in Figs. 1(a) and the 23th order in Figs. 1(b). The
pronounced maximum is at the 15th order in Figs. 1(b). From the
above observations, we find that the interference patterns in the
HHG usually depend on the laser intensity. Moreover, for the large
$R$ case, it becomes difficult for the simple point-emitters model
to predict the interference patterns.

To illuminate the hump and dip structure in Figs. 1(b), we
investigate the physical mechanism for the emission of a harmonic
order  from molecules with large $R$. In Fig. 4, using the same
laser and molecule parameters as in Figs. 1(b), we plot the values
of $S_{2}(\mathbf{p})$ without considering the interference term
Eq. (4) (the red curves with the hollow symbol) and
$S(\mathbf{p})$ with considering the interference term Eq. (4)
(the black curves with the solid symbol) as functions of the
electron momentum $\mathbf{p}$ for certain individual harmonic
orders. Here
\begin{eqnarray}
S_{2}(\mathbf{p})=J_{n}(\frac{-\mathbf{p}\cdot{\mathbf{A}_{0}}}{{\omega}})J_{m}(\frac{\mathbf{p}\cdot{\mathbf{A}_{0}}}{{\omega}}),
\end{eqnarray}
and
\begin{eqnarray}
S(\mathbf{p})=\langle 0|\mathbf{p}\rangle
(\mathbf{p}\cdot\hat{\mathbf{e}})^{2}\langle \mathbf{p}|0\rangle
J_{n}(\frac{-\mathbf{p}\cdot{\mathbf{A}_{0}}}{{\omega}})J_{m}(\frac{\mathbf{p}\cdot{\mathbf{A}_{0}}}{{\omega}}).
\end{eqnarray}
And only the contributions from positive momenta are shown. For
clarity, we use the dotted arrows to indicate those momenta, which
correspond to the first several minimal  extrema  of
$S_{1}(\mathbf{p})$ in  Fig. 3. The relevant momentum values are
labelled above the arrows. We use the horizontal arrows in Figs.
4(a) to indicate several maximal extrema of $S_{2}(\mathbf{p})$
(denoted as $peak_{1}$ and $peak_{2}$).

From Eq. (6), the contributions to single peak of HHG consist of
two parts, i.e., the interference term $S_{1}(\mathbf{p})$ and
Bessel function term $S_{2}(\mathbf{p})$. For the $23th$ order
harmonic at which the harmonic dip appears in Figs. 1(d), without
considering  the interference term $S_{1}(\mathbf{p})$, the
primary contributions of $S_{2}(\mathbf{p})$ come from those
momenta around  the values of $p=0.77$ a.u. and $p=1.37$ a.u.
corresponding to two of the minimal extrema of
$S_{1}(\mathbf{p})$, as indicated by the red curve with the hollow
symbol in Figs. 4(a). When considering the interference term
$S_{1}(\mathbf{p})$, the contributions from those  momenta are
strongly suppressed, as indicated by the black curve with the
solid symbol in Figs. 4(a). In Figs. 4(a) those momenta around the
value of $p=1.02$ a.u. give positive and negative contributions to
the amplitude of the $23$th order harmonic. Because the positive
and negative values will cancel each other so the sum of these
contributions is small.

For other harmonic orders, for example, the $13$th order, around
which the harmonic hump appears in Figs. 1(d). The red curve with
the hollow symbol in Figs. 4(b) shows that the primary
contributions of $S_{2}(\mathbf{p})$ to the harmonic order come
from those momenta around the value of $p=1.02$ a.u., different
from $p=0.77$ a.u. and $p=1.37$ a.u. of the $23$th order case. The
black curve with the solid symbol in Figs. 4(b) also shows that
the suppression effect from the interference term
$S_{1}(\mathbf{p})$ is weaker than that of the $23$th order case.
The above observations indicate that the interference term
$S_1(\mathbf{p})$ is mainly responsible for the formation of the
hump and dip in Figs. 1(b).

Fig. 4  clearly shows that the formation of single order harmonic
is closely related to  the recombination electrons with many
different  momenta $\mathbf{p}$. The contribution from the
recombination electron at one certain momentum $\mathbf{p}$ is
weighted by the corresponding probability amplitude
$S(\mathbf{p})=S_{1}(\mathbf{p})S_{2}(\mathbf{p})$. In addition,
from Figs. 4(a) it can also be concluded that the interference
minimum in the harmonic spectrum could  appear at the harmonic
order at which the primary contributions from the term
$S_{2}(\mathbf{p})$ are strongly suppressed due to the destructive
interference from the term $S_{1}(\mathbf{p})$. Since the
probability amplitude $S(\mathbf{p})$ depends on the field
intensity, the position of the harmonic hump or dip depends on the
intensity too.

The interference patterns in HHG for molecules with large $R$ are
much more complicated than that for molecules with small $R$. The
adjacent  maximal and minimal extremum points of
$S_{1}(\mathbf{p})$ correspond to constructive and destructive
interferences. They are so close as shown in Fig. 3, that the
constructive and destructive interference extrema  in the
plottings of $S(\mathbf{p})$ could be entangled. This makes the
interference effects on HHG very sensitive to the field intensity.
In this case,  the simple point-emitter model is not available.
Our model could provide a good description for the HHG of diatomic
molecules even for the case of  large internuclear separation.

\section{Conclusion}
In conclusion, using a  SFA model that considers Coulomb potential
modification on the continuum wavefunctions, we have analytically
and numerically investigated the HHG from diatomic molecules with
large internuclear separation. The harmonic spectra obtained by
the SFA  model agree with the numerical simulations. Our model
calculation reveals that the two-center interference as well as
the interference between different recombination electron
trajectories are responsible for the unusual enhanced or
suppressed harmonic yield at a certain order, and these
interference effects depend on the laser parameters such as
intensity

\section{Acknowledgements}
This work is supported by 973 research program No. 2006CB806000,
NNSF(No.10725521).

\end{document}